\documentclass[12pt]{article}

\usepackage{hyperref}
\usepackage{units}
\usepackage{epsfig}
\usepackage{pifont}
\usepackage{amsmath,amsfonts,amssymb,amsthm,mathrsfs}
\usepackage{graphicx,chngcntr,slashed}
\usepackage{cancel}
\usepackage{pstricks}
\usepackage{afterpage}
\usepackage{braket}
\usepackage{caption}
\usepackage{subcaption}
\usepackage{xcolor}
\usepackage{makecell}
\usepackage{physics}
\usepackage{multirow}
\usepackage{float}

\newcommand{\pp}{p^{\prime}}
\newcommand{\be}{\begin{equation}}
\newcommand{\ee}{\end{equation}}
\newcommand{\bea}{\begin{eqnarray}}
\newcommand{\eea}{\end{eqnarray}}

\def\pp#1{\left(#1\right)}
\def\bb#1{\left[#1\right]}

\def\d#1{\mathop{{\rm d}#1}}

\makeatletter
\def\section{\@startsection {section}{1}{\z@}{+3.0ex plus +1ex minus
  +.2ex}{2.3ex plus .2ex}{\large\bf\boldmath}}
\def\subsection{\@startsection{subsection}{2}{\z@}{+2.5ex plus +1ex
minus +.2ex}{1.5ex plus .2ex}{\normalsize\bf\boldmath}}
\def\subsubsection{\@startsection{subsubsection}{3}{\z@}{+3.25ex plus
 +1ex minus +.2ex}{1.5ex plus .2ex}{\normalsize\it}}

\graphicspath{{.}{plots/}}

\begin{document}
\thispagestyle{empty}

\def\thefootnote{\fnsymbol{footnote}}

\begin{flushright}
\end{flushright}

\vspace{1cm}

\begin{center}

{\Large {\bf Transverse spin asymmetries and the electron Yukawa coupling at an FCC-ee}}
\\[3.5em]
%
%


{\large
Radja~Boughezal$^1$, Frank~Petriello$^{1,2}$ and Ka{\u{g}}an~\c{S}im\c{s}ek$^2$
}

\vspace*{1cm}

{\sl
$^1$ HEP Division, Argonne National Laboratory, Argonne, Illinois 60439, USA \\[1ex]
$^2$ Department of Physics \& Astronomy, Northwestern University,\\ Evanston, Illinois 60208, USA
}

\end{center}


\begin{abstract}
    
  We show that measurements of single transverse-spin asymmetries at an $e^+e^-$ collider can enhance the sensitivity to the electron Yukawa coupling, possibly enabling observation of the Standard Model value for this quantity. We demonstrate that the significance in both the $b\bar{b}$ and semi-leptonic $WW$ final states can be enhanced by factors of up to three compared to inclusive cross section determinations of this coupling for transversely-polarized electrons. If the positrons can be simultaneously longitudinally polarized, even at the level of 30\%, the significance can be enhanced by a factor of five or more. The method utilizes quantum interference between the Higgs signal and the continuum background and is also applicable in other $WW$ and $ZZ$ final states.
  
\end{abstract}

\setcounter{page}{0}
\setcounter{footnote}{0}


\section{Introduction}

Probing the properties of the Higgs boson is of upmost importance to the future high energy physics program. The role of the Higgs field in the Standard Model (SM) is to give masses to elementary particles through its couplings to these states. As the masses of the SM particles and therefore the couplings of these particles to the Higgs vary by many orders of magnitude, it is imperative to test whether the single Higgs boson in the SM is indeed fully responsible for this enormous spread of particle masses. This program was started at the LHC, and today we have measurements of the Higgs couplings to gauge bosons, third-generation fermions and several lighter fermions~\cite{ATLAS:2022vkf,CMS:2022dwd}. 

The smallest of the SM Higgs couplings is to the electron, which in the SM has the value $y_e^{SM} = \sqrt{2} m_e/v = 2.9 \times 10^{-6}$, with $v$ the Higgs vacuum expectation value. The current limit on this quantity at the LHC is $|y_e| \leq 260 |y_e^{SM}|$ at 95\% CL, coming from searches in the Drell-Yan channel~\cite{CMS:2014dqm,ATLAS:2019old}. If this result improves as the square root of the accumulated integrated luminosity, than a limit $|y_e| \leq 120 |y_e^{SM}|$ will be possible with 3 ab$^{-1}$ of integrated luminosity~\cite{Cepeda:2019klc}. This result can be improved at a high-luminosity $e^+e^-$ collider operating at the Higgs mass, $\sqrt{s}=125$ GeV, such as the Future Circular Collider (FCC) with $e^+e^-$ beams~\cite{FCC:2018evy}. This has been the subject of several studies~\cite{Greco:2016izi,Dery:2017axi,dEnterria:2021xij}. Even with this resonant enhancement the $s$-channel Higgs production cross section that is sensitive to the electron Yukawa coupling  reaches only about 1 fb due to the $y_e^2$ suppression, with backgrounds several orders of magnitude higher. The most complete study of the electron Yukawa coupling at the FCC found that statistically significant results can be obtained in the $H \to gg$, $H \to WW$, and $H\to ZZ$ channels with a multivariate analysis~\cite{dEnterria:2021xij}. A combination of all Higgs final states assuming 10 ab$^{-1}$ of integrated luminosity led to the limit $|y_e| \leq 1.6 |y_e^{SM}|$. Numerous experimental challenges must be overcome to achieve this result, including an ultra-precise measurement of both $M_H$ and $\Gamma_H$, as well as exquisite control over the beam energy spread to avoid reduction of the signal cross section.

In this paper we show that it may be possible to improve upon the determination of the electron Yukawa coupling using transverse polarization of the initial-state electron beam at an FCC. The basic idea is that single transverse spin asymmetries are chiral quantities that are necessarily  proportional to the electron mass. This is well known from studies of these quantities in low-energy deep-inelastic scattering~\cite{Metz:2006pe}. This provides a chiral suppression of the continuum background, reducing it to the level of the Higgs signal. The asymmetry arises from quantum interference between the Higgs exchange and other diagrammatic contributions, leading to an asymmetry linearly proportional to the electron Yukawa coupling, rather than quadratically proportional to the electron mass like the inclusive cross section. This partially ameliorates the strong suppression provided by the small electron Yukawa coupling. The signal has a distinct dependence on the azimuthal angle defined by the final-state particles and the transverse-spin direction, allowing background contributions to be removed by appropriate angular weighting of the asymmetry numerator. Further improvement is possible if the positron beam can be simultaneously longitudinally polarized even at the level of 30\%. We use the statistical significance, the number of signal events over the square root of the number of background events, as our figure-of-merit. We find that the significance of the $b\bar{b}$ channel can be improved by a factor of five compared to the inclusive cross section analysis, while the semi-leptonic $WW$ channel can be improved by a factor of six. In the case of the semi-leptonic $WW$ final state the significance reaches nearly three, and could lead to the observation of the SM electron Yukawa coupling. The simultaneous need to reach high integrated luminosity, beam polarization and small beam energy spread leads to stringent requirements on accelerator and detector performance. The study of whether the FCC is capable of the required level of performances is a subject under investigation~\cite{Blondel:2019jmp}. However, if possible, this may facilitate the measurement of one of the last remaining parameters of the SM.

This paper is organized as follows. We review the properties of transverse spin asymmetries in Section~\ref{sec:transpin}, with a focus on showing the properties that make them helpful for the determination of the electron Yukawa coupling. We present numerical results in Section~\ref{sec:numerics}. We present there also our calculational formalism that accounts for initial-state radiation and beam-energy spread, and explain the cuts we use to enhance the value of the asymmetries. We show the significance achievable for both the $b\bar{b}$ and semi-leptonic $WW$ final states as a function of various experimental parameters. We conclude in Section~\ref{sec:conc}.

\section{Review of transverse spin asymmetries}
\label{sec:transpin}

We review here the calculation of transverse spin asymmetries in $e^-e^+$ collisions.  We assume that the electron is transversely polarized. Taking both initial beams along the $\hat{z}$-axis, we can write the transverse spin vector of the electron as
\be
    S_T^{\mu} = \left(0,\text{cos}(\phi),\text{sin}(\phi),0 \right)
\ee
where $\phi$ denotes the angle between the transverse spin and the direction of an outgoing particle in the transverse plane.  In our study we will consider the following two processes:
\begin{gather}
    e^- (p_a) + e^+ (p_b) \to b (p_1) + \bar b (p_2), \\ 
    e^- (p_a) + e^+ (p_b) \to W (p_{12}) + W (p_{34}) \to \ell (p_1) + \nu (p_2) + j (p_3) + j (p_4).
\end{gather}
In the $b\bar{b}$ case the outgoing particle chosen to define $\phi$  is the $b$-quark, while in the $WW$ case it is the leptonically decaying-$W$ boson. We will consider several asymmetries including ones where the positron is longitudinally polarized. Since we are interested in terms in the amplitude proportional to the electron mass, we must be careful in our definition of the longitudinal polarization vector. We define it in terms of the initial momenta $p_a$ and $p_b$ as $s_b = c_a p_a + c_b p_b$, where
\begin{gather}
    c_a = - {2 m_e \over E \sqrt{E^2 - 4 m_e^2}},\quad 
    c_b = - {2 m_e^2 - E^2 \over m_e E \sqrt{E^2 - 4 m_e^2}}.
\end{gather}
This leads to the following conditions being satisfied, as required: $p_a^2 = p_b^2 = m_e^2$, $s_a^2 = s_b^2 = -1$, $p_a \cdot s_a = p_b \cdot s_b = 0$, $s_a \cdot s_b = 0$, and $p_a \cdot s_b = {E \sqrt{E^2 - 4m_e^2} \over 2m_e}$. Here, $E$ denotes the center-of-mass energy. We note that the expression above for $s_b$ is equivalent to the usual definition $s_b = {1 \over m_e} (|\vec p_b|, E_b \hat{p}_b)$. With these expressions we can form projection operators which allow us to easily polarize the initial-state particles in our calculation:
 \begin{gather}
    u_a \bar u_a = (\slashed p_a + m_e) \mathbb P_a, \quad 
    v_b \bar v_b = (\slashed p_b - m_e) \mathbb P_b,
\end{gather}
with
\begin{gather}
    \mathbb P^\pm (\lambda, s) = {1 \pm \lambda \gamma_5 \slashed S_i \over 2}
\end{gather}
and $ \mathbb P^\pm$ chosen to obtain positive/negative spin for the electron, respectively (negative/positive for the positron). For the spin vector $S_i$ we use $S_T$ for the electron and $S_L$ for the positron. We carry out our calculations using {\tt FEYNARTS}~\cite{Hahn:2000jm} and {\tt FEYNCALC}~\cite{Shtabovenko:2020gxv}, and cross-check these results with in-house codes. For the $b\bar{b}$ process, we keep the mass of the $b$ quark to first order. This is necessary as the Higgs signal is proportional to this quantity. For the $WW$ process we take the outgoing fermions to be massless. For both processes we keep the mass of the initial-state electrons to first order. This procedure allows us to extract the overall electron-mass dependence of the amplitude without keeping numerically negligible quadratic and other higher-order mass terms. 

As mentioned we consider several different asymmetries in our analysis in order to determine what can be gained from longitudinal polarization as well. We write these asymmetries generically as  $A = N/D$, where $N$ is the difference between cross-section measurements with distinct incoming beam polarizations and $D$ is their sum. We denote the cross sections by $\sigma^{\lambda_a \lambda_b}$ where $\lambda_a$ and $\lambda_b$ can take on the values $+1$, 0, and $-1$. The first superscript refers to the transverse polarization of the electron while the second refers to the longitudinal polarization of the positron. Although not explicitly written we will consider these cross sections as functions of the final-state kinematics and allow cuts to be imposed on both numerator and denominator, and we will later introduce a weight function to project out the terms proportional to the electron Yukawa coupling in the numerator. We consider the polarization asymmetries constructed below.
\begin{itemize}
    \item The double-polarization asymmetry (DP):
    \begin{gather}
        N = \frac14 (\sigma^{++} - \sigma^{+-} - \sigma^{-+} + \sigma^{--}) \\
        D = \frac14 (\sigma^{++} + \sigma^{+-} + \sigma^{-+} + \sigma^{--}) 
    \end{gather}
    \item The single-polarization asymmetry with an unpolarized positron beam, $\lambda_b = 0$ (${\rm SP}^0$):
    \begin{gather}
        N = \frac12 (\sigma^{+0} - \sigma^{-0}) \\ 
        D = \frac12 (\sigma^{+0} + \sigma^{-0})
    \end{gather}
    \item The single-polarization asymmetry with a left-handed positron beam, $\lambda_{b} = +1$ (${\rm SP}^+$):
    \begin{gather}
        N = \frac12 (\sigma^{++} - \sigma^{-+}) \\
        D = \frac12 (\sigma^{++} + \sigma^{-+})
    \end{gather}
    \item The single-polarization asymmetry with a right-handed positron beam, $\lambda_{b} = -1$ (${\rm SP}^-$):
    \begin{gather}
        N = \frac12 (\sigma^{+-} - \sigma^{--}) \\
        D = \frac12 (\sigma^{+-} + \sigma^{--})
    \end{gather}
\end{itemize}

Before presenting calculations and numerics it is useful to review the properties of single transverse-spin asymmetries in order to understand why they are helpful in measuring the electron Yukawa coupling. We first note that single transverse-spin asymmetries are linearly proportional to the electron mass. This was pointed out recently in the context of new physics searches in deep-inelastic scattering~\cite{Boughezal:2023ooo} and is well known from previous work on transverse single-spin asymmetries in QCD~\cite{Metz:2006pe}. We will show later that the asymmetries are directly proportional to the electron-Higgs coupling and not the electron mass, which in principle could differ from that inferred from the electron mass in models of new physics. Previous studies of the electron Yukawa coupling at the FCC~\cite{dEnterria:2021xij} were based on measurements of the inclusive cross section on the Higgs resonance peak, which is quadratically proportional to the electron mass. Given the tiny size of the electron-Higgs having a linear dependence on this parameter gives a significant boost to the analysis.

To demonstrate how transverse single-spin asymmetries allow the electron-Higgs coupling to be isolated we briefly review their transformation properties under discrete symmetries. We can show that single transverse-spin asymmetries are odd under the combined transformation of parity and ``naive" time-reversal~\cite{Hagiwara:1982cq} symmetries. Writing a raising operator for a particle with spin $s$ and 3-momentum $\vec{p}$ as $a^{s \dagger}_{\vec{p}}$, we can write these two transformations as
\bea
P (c \,a^{s \dagger}_{\vec{p}}) P^{-1} &=& c \,a^{s \dagger}_{-\vec{p}},\nonumber \\
A (c \,a^{s \dagger}_{\vec{p}}) A^{-1} &=& c \,a^{-s \dagger}_{-\vec{p}},
\eea
where $c$ is a $c$-number. The naive time-reversal operator $A$ differs from the normal time-reversal operator in that it is unitary, and does not conjugate the $c$-number. The generation of an $A$-odd structure requires an imaginary part of the amplitude~\cite{Hagiwara:1982cq}. This requires a loop effect, and the dominant higher-order effect that generates this term is near the Higgs resonance where Dyson resummation introduces the Higgs width and consequently an imaginary part into the propagator. Denoting the momentum of the final-state $b$-quark (for the $b\bar{b}$ process) or leptonically decaying $W$-boson (for the $WW$ process) as $p_f$, we find that only two structures contribute to these asymmetries: $S_T \cdot p_f$ and $\epsilon_{\mu\nu\rho\sigma} p_a^{\mu} p_b^{\nu} p_f^{\rho} S_T^{\sigma}$. It can be straightforwardly checked that the first structure is $P$-odd and $A$-even, while the second is $P$-even and $A$-odd. We can calculate these two structures in terms of the lab-frame polar angle $\theta$ and azimuthal angle $\phi$ as
\bea
S_T \cdot p_f &=&  = \beta_f \frac{E}{2} \text{sin}(\theta) \text{cos}(\phi), \nonumber \\
\epsilon_{\mu\nu\rho\sigma} p_a^{\mu} p_b^{\nu} p_f^{\rho} S_T^{\sigma} &=& -\beta_a \beta_f \frac{{E}^3}{4} \text{sin}(\theta) \text{sin}(\phi),
\eea
where the $\beta_i$ denote the lab-frame velocities of particle $i$ (these would be unity in the massless limit). The angle $\phi$ is defined by the plane formed by $p_f$ and $S_T$. We note that these two structures have different behavior in $\phi$, and can be separated by the appropriate angular weight function.

We show how this works using the $b\bar{b}$ process as an example. At tree level the process proceeds through $s$-channel exchanges of a photon, a $Z$-boson, and the Higgs boson. After squaring diagrams, using the projection operators above and summing over final-state spins, we can write the numerator of the asymmetry in the following schematic form:
\bea
N &=& \frac{1}{2E^2} \int d \text{LIPS} \, {\rm sin}(\phi) \, \left\{ \frac{R_{\gamma\gamma}}{s^2}+\frac{R_{ZZ}}{(s-M_Z^2)^2} + \frac{R_{\gamma Z}}{s(s-M_Z^2)}
    +\frac{R_{\gamma H}(s-M_H^2)}{s[(s-M_H^2)^2+M_H^2\Gamma_H^2]} \right. 
    \nonumber \\ && \left. +\frac{R_{Z H}(s-M_H^2)+I_{ZH}M_H\Gamma_H }{(s-M_Z^2)[(s-M_H^2)^2+M_H^2\Gamma_H^2]} \right\},
\eea
where LIPS denotes the Lorentz-invariant final-state phase space and $s$ the usual Mandelstam invariant. We have included the ${\rm sin}(\phi)$ factor needed for the projection mentioned before. The $R_i$ and $I_i$ respectively denote the real and imaginary parts of the amplitudes after extracting the internal propagators as indicated. We have assumed that we are near the Higgs resonance peak and have introduced the Higgs width into its propagator. We write below explicit expressions for these structures for the DP asymmetry in terms of the various charges and couplings appearing in the SM.
\bea
R_{\gamma\gamma} &=& 96 e^4 Q_e^2 Q_b^2 m_e (S_T \cdot p_f ) (t-u)\nonumber \\
R_{ZZ} &=& 96 m_e (S_T \cdot p_f ) g_Z^4 g_{v}^{e,2} (g_{v}^{b,2}+g_{b}^{b,2})(t-u)+192 m_e (S_T \cdot p_f )g_Z^4 g_{ve}g_{ae}g_{vb}g_{ab} s \nonumber \\
R_{\gamma Z} &=& 192 e^2 g_Z^2 Q_e Q_b m_e (S_T \cdot p_f ) g_{v}^e g_{v}^b (t-u) +96 e^2 g_Z^2 Q_e Q_b m_e (S_T \cdot p_f ) g_{a}^e g_{a}^b s \nonumber \\
R_{\gamma H} &=& -96 e^2 Q_e Q_b y_e y_b (S_T \cdot p_f ) m_b \nonumber \\
R_{ZH} &=& -96 g_Z^2 g_{v}^e g_{v}^b y_e y_b (S_T \cdot p_f ) s \nonumber \\
I_{ZH} &=& -192 g_Z^2 g_{a}^e g_{v}^b y_e y_b \epsilon(p_a,p_b,p_f,S_T).
\eea
We have introduced the abbreviation $\epsilon(p_a,p_b,p_f,S_T) = \epsilon_{\mu\nu\rho\sigma} p_a^{\mu} p_b^{\nu} p_f^{\rho} S_T^{\sigma}$. $s,t,u$ denote the usual Mandelstam variables for $2 \to 2$ scattering. The coupling factors used in this expression are defined in the Appendix. We note several important facts about these formulae.
\begin{itemize}

\item All the $R_i$ and $I_i$ terms are suppressed by the mass of the electron or the electron-Higgs Yukawa coupling.

\item It is straightforward to check that the square of the Higgs exchange diagram is proportional to the factor $S_T \cdot (p_a+p_b) =0$. 

\item The factor containing $I_{ZH}$ is enhanced for $\sqrt{s}=M_H$ since the Higgs propagator becomes proportional to the Higgs width. This term arises from the imaginary part of the interference ${\cal M}_Z \times {\cal M}_H^{*}$, and is proportional to ${\rm sin}(\phi)$ because of the discrete symmetry arguments discussed above. Since this term is linear in the Higgs amplitude it is directly proportional to $y_e$, the Higgs Yukawa coupling. The other structures $R_{\gamma\gamma}$, $R_{ZZ}$, $R_{\gamma Z}$ and $R_{\gamma H}$ are all proportional to ${\rm cos}(\phi)$. The angular weighting with $s_{\phi}$ will remove all but the $I_{ZH}$ term in the cross section.

\end{itemize}

These facts hold for the $WW$ process as well in the limit $E \approx M_H^2$. This case is slightly more complicated due to the $t$-channel neutrino exchange diagram, which leads to structures proportional to ${\rm sin}(\phi)$ and to the electron mass rather than the electron Yukawa coupling. These terms are suppressed by $\Gamma_H/m_h$ near the Higgs resonance. We include the on-shell $WW$ amplitudes in the Appendix in order to illustrate these points. These considerations lead to the following strategy for the determination of the electron-Higgs coupling: form one of the transverse spin asymmetries defined above; weight numerator events by the azimuthal angle factor; go close to the Higgs resonance peak. We note that the angular weight functions $\sin(\varphi)$ and $\pm \sin(\varphi_{12})$ are introduced to the phase space when we form the numerator of the asymmetry of the $b\bar b$ and $WW$ processes, respectively, where the sign of the latter is opposite to the sign of the electric charge of the outgoing lepton, $\ell^\mp$ and $\sin(\varphi_{12})$ refers to the angle associated with the leptonically-decaying $W$ boson. We note that defining the angle $\phi$ requires distinguishing the $b$-initiated jet from the $\bar{b}$-jet. This issue was faced at LEP when measuring the forward-backward asymmetry in the $b\bar{b}$ process. A combination of jet-charge measurements and determinations of the $b$-initiated jet from semi-leptonic $b$ decays were used to facilitate this, resulting in efficiencies around 90\%~\cite{Liebeg:2003}. Since this is connected to the issue of $b$-jet tagging we do not attempt to assign any efficiency in this analysis, and simply assume that the FCC will have equally good performance.

\section{Numerical results}
\label{sec:numerics}

We now present numerical results for the following processes:
\begin{gather}
    e^- (p_a) + e^+ (p_b) \to b (p_1) + \bar b (p_2), \\ 
    e^- (p_a) + e^+ (p_b) \to W (p_{12}) + W (p_{34}) \to \ell (p_1) + \nu (p_2) + j (p_3) + j (p_4).
\end{gather}
We note that the approach described in the previous section can be implemented for other decay processes where the Higgs production diagram interferes with the continuum process. This quantum interference is needed to generate the linear dependence on the electron Yukawa coupling. It can therefore be applied to the fully leptonic and fully hadronic $W$-decay processes. It cannot be applied to the gluonic final state that leads to the highest sensitivity in~\cite{dEnterria:2021xij}, since no continuum gluon pair-production process exists at tree level. There is interference with loop-induced continuum production but we expect that this effect is small.

\subsection{Formalism}

To define the asymmetry and the expected uncertainty in the experimental measurement we let $N_N$ and $N_D$ denote the number of events determined experimentally in the numerator and denominator of the asymmetry under consideration. In the numerator we define $N_N$ after weighting events by ${\rm sin}(\phi)$ in order to project out the term linear in the electron Yukawa couplings.  The experimental reconstruction of the asymmetry is given by
\begin{gather}
    A^{\rm exp} = {1 \over P_{e^-} P_{e^+}} {N_N \over N_D}
\end{gather}
when both incoming beams are polarized and
\begin{gather}
    A^{\rm exp} = {1 \over P_{e^-}} {N_N \over N_D}
\end{gather}
when only the electron beam is polarized. Here, $P_{e^\mp}$ are the electron/positron beam polarizations at a future FCC. We assume that the electron is transversely polarized and that any positron polarization is longitudinal. Unless stated otherwise we assume the following numerical values for each parameter:
\be
P_{e^-} = 80\%, \;\;\; P_{e^+} = 30\%.
\ee
In the limit of small asymmetry, the error in $A^{\rm exp}$ is given by
\begin{gather}
    \delta A^{\rm exp} = {\delta P_{e^-} \over P_{e^-}} A^{\rm exp} \oplus {\delta P_{e^+} \over P_{e^+}} A^{\rm exp} \oplus {1 \over P_{e^-} P_{e^+}} {1 \over \sqrt{N_D}} \label{dA_pol}
\end{gather}
when both beams are polarized and
\begin{gather}
    \delta A^{\rm exp} = {\delta P_{e^-} \over P_{e^-}} A^{\rm exp} \oplus {1 \over P_{e^-}} {1 \over \sqrt{N_D}} \label{dA_upol}
\end{gather}
when the positron beam is unpolarized. We have denoted a sum in quadrature over the statistical uncertainties and the polarization uncertainties. We do not consider other sources of potential experimental uncertainties associated with reconstruction of final-state objects, consistent with the analysis of ~\cite{dEnterria:2021xij}. Future detectors at an FCC are expected to have excellent resolution and reconstruction capabilities~\cite{Azzi:2021ylt}, and we believe that this is a reasonable assumption. The polarization uncertainties are also expected to be small at a future FCC complex. We assume 3\% uncertainties here but in practice only the statistical uncertainties are relevant for this analysis. Our figure-of-merit is the significance, defined as $\mathcal S = A^{\rm exp} / \delta A^{\rm exp}$. Unless stated otherwise we assume an integrated luminosity of $L = 10\ {\rm ab}^{-1}$, consistent with the study of~\cite{dEnterria:2021xij}.

To obtain a realistic estimate for the possible significance obtainable at an FCC, especially for an analysis that requires tuning the beam collision energy very close to the Higgs resonance, we must account for both beam energy spread and initial-state radiation (ISR). We implement these two effects into our cross section estimates using the following convolution:
\begin{gather}
    \sigma (E_{\rm coll}) = \int_{-\infty}^{\infty} \d{\hat{E}} \ {\d{\mathcal L (E_{\rm coll}, \hat{E}, \delta)} \over \d{\hat{E}}} \int_0^1 \d{x} \  f(x, \hat{E}) \sigma(\sqrt x \hat{E}), \label{sig_bsisr}
\end{gather}
where the beam spread is characterized by a relativistic Voigtian function~\cite{Kycia:2017gjn}
\begin{gather}
    {\d{\mathcal L (E_{\rm coll}, \hat{E}, \delta)} \over \d{\hat{E}}} = {1 \over \sqrt{2\pi \delta^2}} \exp\bb{- {(\hat{E} - E_{\rm coll})^2 \over 2 \delta^2}}.
\end{gather}
Here, $E_{\rm coll}$ is the collider energy and $\delta$ is the c.m. energy spread. By default we assume the value $\delta=4.1$ MeV, the Higgs width in the SM~\cite{dEnterria:2021xij}, although we study the dependence of the significance of our result on this parameter toward the end of this section. We use the Jadach-Ward-Was ISR function~\cite{Jadach:2015cwa,Jadach:2000ir} used in previous studies \cite{Greco:2016izi,deBlas:2022aow}:
\begin{gather}
    f(x, \hat{E}) = \exp\bb{{\beta_e \over 4} + {\alpha \over \pi} \pp{-\frac12 + {\pi^2 \over 3}}} {\exp(-\gamma \beta_e) \over \Gamma(1 + \beta_e)} \beta_e (1-x)^{\beta_e - 1} \bb{1 + {\beta_e \over 2} - \frac12 (1-x)^2},
\end{gather}
where $\gamma$ is the Euler-Mascheroni constant and for this formula we set 
\begin{gather}
    \beta_e = {2\alpha \over \pi} \bb{\ln\pp{\hat{E}^2 \over m_e^2} - 1}.
\end{gather}

\subsection{Analysis of the $b\bar{b}$ channel}

We begin by discussing the $b\bar{b}$ final state. In the analysis of~\cite{dEnterria:2021xij} based on the inclusive cross section a significance of $\mathcal S = 0.12$ was found. The limiting factor in the analysis was the continuum $b$-jet irreducible background. Multi-variate techniques were unable to improve upon the rejection of these events. We focus here on the continuum $b\bar{b}$ background, consistent with this previous study. In our analysis we begin by choosing the following basic cuts on our final state: $m_{\rm inv, cut} = 120  \,{\rm GeV}$ and $5^o < \theta < 175^o$. We assume a reconstruction efficiency of the $b\bar{b}$ system of 80\%, following~\cite{dEnterria:2021xij}. For the polarization asymmetries defined earlier we find the values for the significance shown in the first column of Table~\ref{tab:bb}. The reference value in Table~\ref{tab:bb} refers to the significance obtained using the inclusive cross section instead of a polarization asymmetry. Our result of $\mathcal S = 0.11$ is in good agreement with the result of~\cite{dEnterria:2021xij}, which provides a check of our analysis. As an additional check of our result, we note that the fully inclusive cross section we obtain for the continuum background including both beam-spread and ISR effects is 15 pb, in reasonable agreement with the 19 pb assuming only beam-spread effects found in~\cite{dEnterria:2021xij}

\begin{table}
    \centering
    \begin{tabular}{|c||c|c|c|}
        \hline 
        \hline 
        Observable & Basic cuts & Enhanced cuts & $\theta_{\rm cut}$ [\%]\\
        \hline
        ${\rm DP}$ & 0.27  & 0.41 & 39\\ 
        ${\rm SP}^0$ & 0.19  & 0.30 & 33 \\
        ${\rm SP}^+$ & 0.11  & 0.17 & 44 \\
        ${\rm SP}^-$ & 0.37  & 0.58 & 39 \\
        \hline 
        Reference & 0.11 &$-$ & $-$  \\
        \hline\hline 
    \end{tabular}
  \caption{The significance estimates for the double-polarization and various single-polarization asymmetries in the $b\bar{b}$ channel, as well as for the inclusive cross section (denoted as ``reference") obtained with an integrated luminosity of $L = 10\ {\rm ab}^{-1}$, and beam polarization values of $P_{e^-} = 80\%$, $P_{e^+} = 30\%$. The basic cuts refer to $m_{\rm inv, cut} = 120  \,{\rm GeV}$ and $5^o < \theta < 175^o$, while the enhanced cuts refer to  $m_{\rm inv, cut} = m_h - 10 \,{\rm MeV}$ and the indicated percentages of the interval $[0^o,180^o]$ removed. } 
    \label{tab:bb}
\end{table}

We note that the single-polarization asymmetry ${\rm SP}^0$ where only the electron is transversely polarized gives a significance of $\mathcal S=0.19$ compared to the inclusive cross section result of $\mathcal S = 0.11$, showing the gain possible using transverse polarization. Further longitudinal polarization of the positron results in a significance of $\mathcal S = 0.37$, over three times larger than the reference value. We will discuss in detail the reason for this significant improvement when we discuss the $WW$ channel. 

\begin{figure}[htbp]
    \centering
    \includegraphics[width=.45\textwidth]{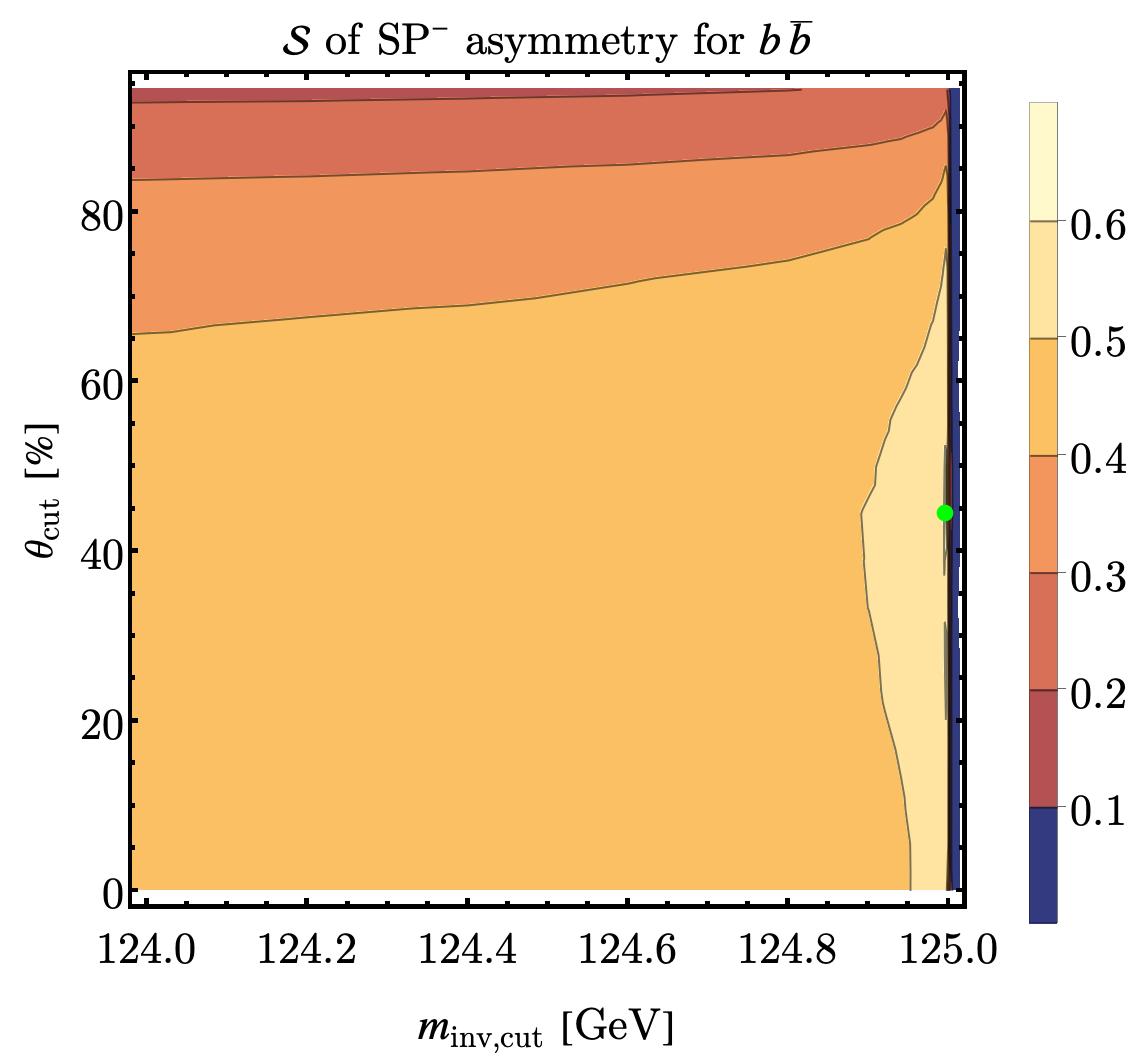}
    \includegraphics[width=.45\textwidth]{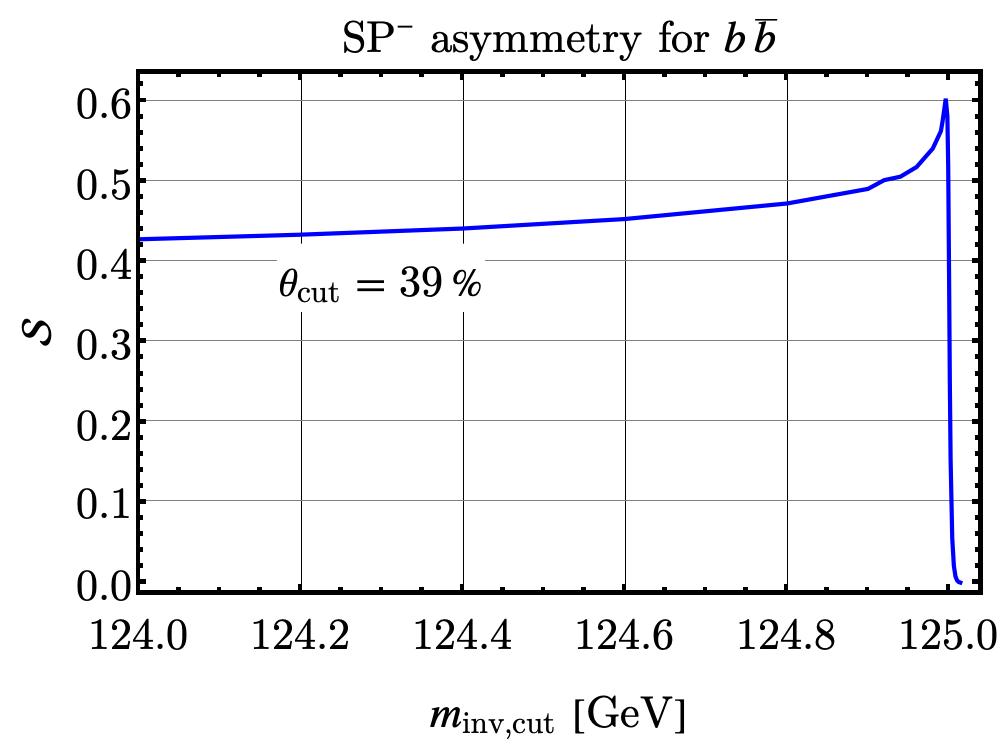}
    \caption{The dependence of the sensitivity estimate for the ${\rm SP}^-$ asymmetry on the invariant-mass and polar-angle cuts for the $b\bar b$ process. The left panel shows the dependence of the significance on the polar-angle cut for the basic invariant-mass cut of 120 GeV. We present this in terms of the percentage of the phase-space removed by symmetric cuts on the left and right parts of the interval $[0^o, 180^o]$. The right panel shows the dependence of the significance on the invariant mass cut for the best theta-cut value. The green dots denote the maximum significances found by imposing these cuts. }
    \label{fig:bb_thetacut_minvcut}
\end{figure}

We can improve upon the significance of our results by imposing more restrictive cuts on the final-state phase space. Although further cuts reduce the number of events and therefore increase the statistical uncertainty, the statistical uncertainty only scales as the square root of the number of events. If the introduced cuts enhance the asymmetry more quickly than this rate we can improve the significance. We note that the Higgs is a spin-0 particle while the continuum background is mediated by the spin-1 photon and $Z$-boson. This leads to an angular distribution for the numerator peaked near central polar angles while the denominator is peaked toward forward polar angles. This suggests that a more stringent cut on the polar angle could enhance the asymmetry. We also note that a stronger cut on the final-state invariant mass could increase the asymmetry as well, since it focuses on the Higgs peak where the asymmetry is generated and reduces the continuum $b\bar{b}$ background. Neither of these cuts depend on the azimuthal angle $\phi$ and consequently do not affect that angular weighting needed to form the asymmetry. We study the dependence on these more stringent cuts in Fig.~\ref{fig:bb_thetacut_minvcut}. The left panel shows the dependence of the significance on the polar-angle cut for the basic invariant-mass cut of 120 GeV in terms of the percentage of the polar-angle interval $[0^o, 180^o]$ removed. The right panel shows the dependence on the invariant mass cut for the optimal polar angle cut. In both cases there is a trade-off between stronger cuts that enhance the asymmetry but simultaneously increase the statistical uncertainty. The best-case value of the theta cut removes 39\% of the available phase space for the ${\rm SP}^-$ asymmetry, and slightly different values for the other asymmetries. For the invariant mass cut we focus on the region $m_{\rm inv}=[124,125]$ GeV due to the excellent expected resolution at a future FCC. When we get too close to the Higgs mass we begin to cut into the signal region, reducing the sensitivity. We focus on the significance possible with the cut $m_{\rm inv,cut} = m_h - 10 \, {\rm MeV}$, which is very close to the Higgs resonance peak. Whether such a stringent cut is possible depends on the precision achievable with FCC detectors. 

The results for the significances using these enhanced cuts are shown in Table~\ref{tab:bb}. For the case of only a transversely polarized electron the significance for the asymmetry ${\rm SP}^0$ reaches $\mathcal S = 0.30$, nearly a factor of three larger than the reference value obtained using the inclusive cross section. For the best-case polarization choice ${\rm SP}^-$ we estimate that the significance can reach  $\mathcal S = 0.58$, a factor of five greater than the reference value. We stress that this will depend on the possible resolution achievable for the final-state invariant mass cuts. However, we note from  Fig.~\ref{fig:bb_thetacut_minvcut} that even for an invariant mass cut of $m_{\rm inv,cut} = m_h - 1 \, {\rm GeV}$ that a significance of $\mathcal S > 0.40$ can be achieved.

\subsection{Analysis of the $WW$ channel}

We now discuss the $WW$ final state. We focus on the semi-leptonic final state for simplicity and as a demonstration of the measurement. This technique is applicable to the other $WW$ final states as well. In the analysis of~\cite{dEnterria:2021xij} a boosted decision tree (BDT) was used to remove the reducible continuum light-quark background. The authors of this analysis provide a set of simple cuts that approximately reproduce the results of the BDT:
\begin{gather}
    \ E_{j_1, j_2} < 52, 45 \ {\rm GeV}, \ E_\ell > 10 \ {\rm GeV}, \ E_{\rm miss} > 20 \ {\rm GeV}, \ m_{l\nu} > 12 \ {\rm GeV}
    \label{eq:WWbasic}
\end{gather}
where $j_1, j_2$ refer to the jets from the hadronically-decaying $W$-boson, $E_{\rm miss}$ is the missing energy associated with the neutrino from the leptonically-decaying $W$-boson, and $m_{l\nu}$ is the invariant mass of the leptonically-decaying $W$-boson. We incorporate these cuts in our analysis. It can be checked straightforwardly that these cuts do not depend on the azimuthal angle used to define the asymmetry considered here, which allows us to avoid the complication of not integrating over the full angular phase space needed to project out the electron Yukawa coupling contribution. We note that the cross-section value for the irreducible background process  $e^-e^+ \to WW \to \ell \nu jj$  including beam spread and ISR effects is 15 fb, to be compared to 23 fb obtained without ISR in~\cite{dEnterria:2021xij}. We further check our result by analytically integrating the final state over the leptonic phase space and comparing the result to a direct calculation of the asymmetry using the on-shell amplitudes presented in the Appendix and find complete agreement between the two results.

The analysis of the semi-leptonic $WW$ channel proceeds similarly to the $b\bar{b}$ channel. To form the asymmetry we use $\phi_{12}$, the angle between the two reconstructed $W$-bosons. We note that the analysis is slightly more complicated due to the presence of the $t$-channel neutrino exchange diagram contributing to the $WW$ process. The projection of the electron Yukawa term using ${\rm sin}(\phi)$ works only for invariant masses near the Higgs mass, unlike for the $b\bar{b}$ case where the projection works for arbitrary invariant mass. Nevertheless if we focus on the Higgs resonance region our angular-weighted asymmetry isolates the Higgs Yukawa term up to corrections suppressed by the Higgs width over mass ratio. We make these statements more explicit using the simpler analytic formulae for the on-shell $WW$ amplitudes in the Appendix. 

We begin by studying the basic cuts defined above in Eq.~(\ref{eq:WWbasic}) together with the invariant mass cut $m_{\rm inv,cut} =120$ GeV that isolates the Higgs resonance peak. Our results combine both the positively and negatively charged lepton channels, and include all light generations of leptons and quarks. The results are shown in Table~\ref{tab:WW}. The reference value shown in the table comes from an analysis in terms of the inclusive cross section. Our significance of $\mathcal S = 0.45$ is comparable to but slightly worse than the result $\mathcal S = 0.53$ found in~\cite{dEnterria:2021xij}. We expect that this difference is due to the use of simple cuts in our analysis as compared to the full-fledged boosted decision tree analysis used there. We see that for the polarization choice ${\rm SP}^+$ a significance of $\mathcal S=2.0$ can be reached, nearly a factor of five greater than what we find using the inclusive cross section.

\begin{table}
    \centering
    \begin{tabular}{|c||c|c|c|}
        \hline 
        \hline 
        Observable & Basic cuts & Enhanced cuts & $\theta_{\rm cut}$ [\%]\\
        \hline
        ${\rm DP}$ & 0.31  & 0.44 & 22\\ 
        ${\rm SP}^0$ & 0.47  & 0.80 & 44 \\
        ${\rm SP}^+$ & 2.0  & 2.9 & 28 \\
        ${\rm SP}^-$ & 0.12  & 0.22 & 67 \\
        \hline 
        Reference & 0.45 &$-$ & $-$  \\
        \hline\hline 
    \end{tabular}
  \caption{The significance estimates for the double-polarization and various single-polarization asymmetries in the $WW \to l\nu jj$ channel, as well as for the inclusive cross section (denoted as ``reference") obtained with an integrated luminosity of $L = 10\ {\rm ab}^{-1}$, and beam polarization values of $P_{e^-} = 80\%$, $P_{e^+} = 30\%$. The basic cuts refer to $m_{\rm inv, cut} = 120  \,{\rm GeV}$ and  those described in Eq.~(\ref{eq:WWbasic}), while the enhanced cuts refer to  $m_{\rm inv, cut} = m_h - 10 \,{\rm MeV}$ and the indicated percentages of the interval $[0^o,180^o]$ removed. } 
    \label{tab:WW}
\end{table}

\begin{figure}[htbp]
    \centering
    \includegraphics[width=.45\textwidth]{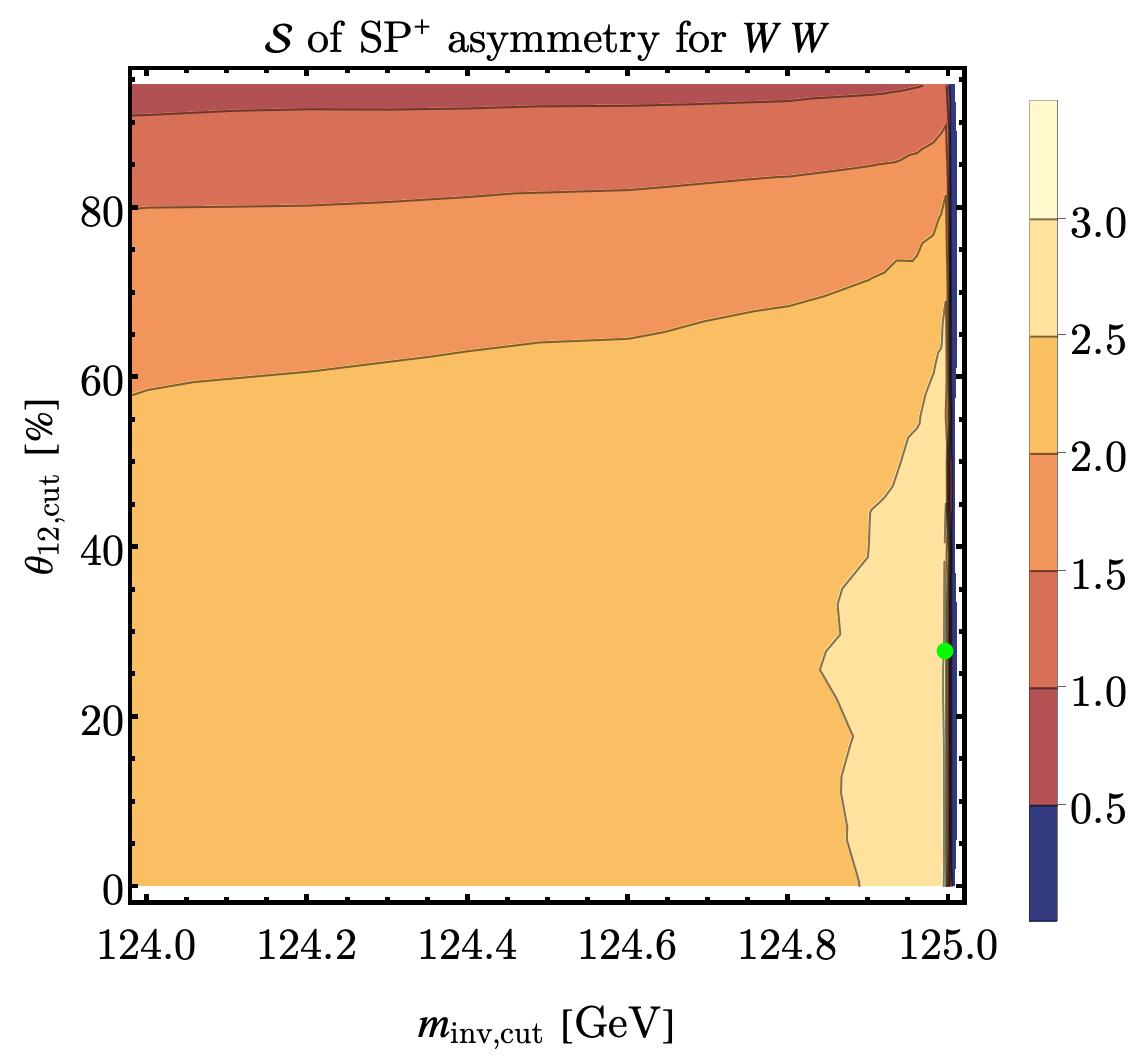}
    \includegraphics[width=.45\textwidth]{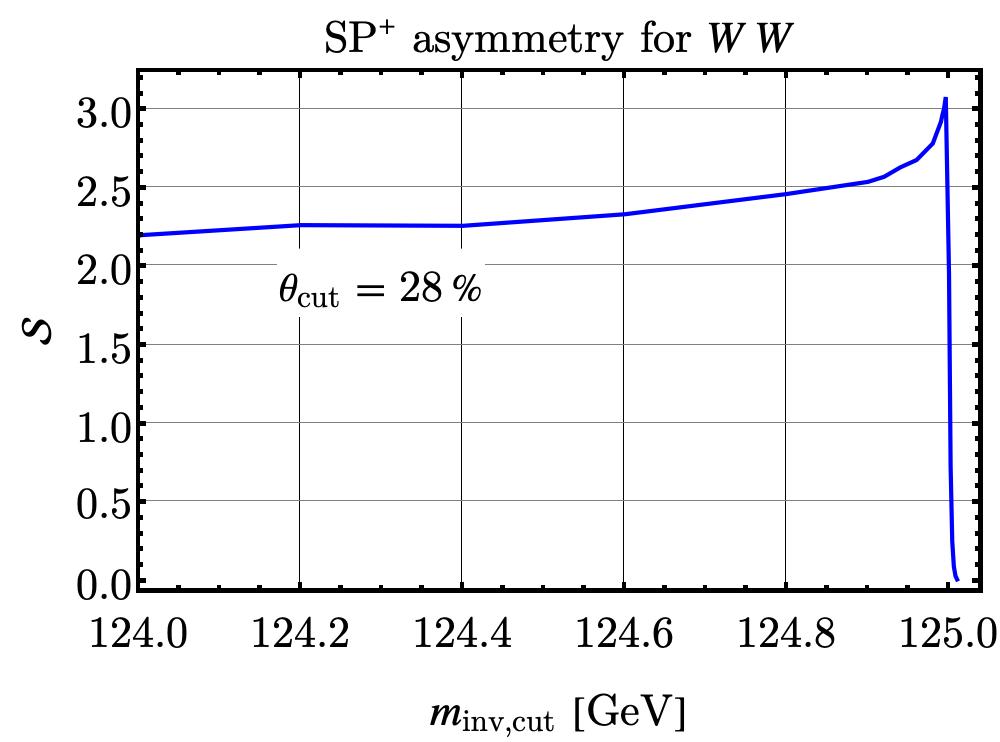}
    \caption{The dependence of the sensitivity estimate of the ${\rm SP}^-$ asymmetry on the invariant-mass and polar-angle cuts for the $b\bar b$ process. The left panel shows the dependence of the significance on the polar-angle cut for the basic invariant-mass cut of 120 GeV. We present this in terms of the percentage of the phase-space removed by symmetric cuts on the left and right parts of the interval $[0^o, 180^o]$. The right panel shows the dependence of the significance on the invariant mass cut for the best theta-cut value. The green dots denote the maximum significances found by imposing these cuts. }
    \label{fig:WW_thetacut_minvcut}
\end{figure}

 We can again improve on the significance by further cuts on the polar angle and invariant mass, following the same logic as for the $b\bar{b}$ channel. The dependence on the polar-angle cut and the invariant-mass cut is shown on Fig.~\ref{fig:WW_thetacut_minvcut} for the 
${\rm SP}^+$ asymmetry, and the significances with the enhanced cuts that maximize the asymmetry are shown in Table~\ref{tab:WW}. If an invariant mass cut of  $m_{\rm inv, cut} = m_h - 10 \,{\rm MeV}$ can be achieved the significance can reach $\mathcal S =  2.9$, nearly a factor of six greater than the reference value. Even a value $m_{\rm inv, cut} = m_h - 1 \,{\rm GeV}$ can provide a significance of approximately $\mathcal S =  2.2$. These values are close to the value $\mathcal S=3$ needed to claim observation of the SM electron Yukawa coupling.

\begin{figure}[H]
    \centering
    \includegraphics[width=.4875\textwidth]{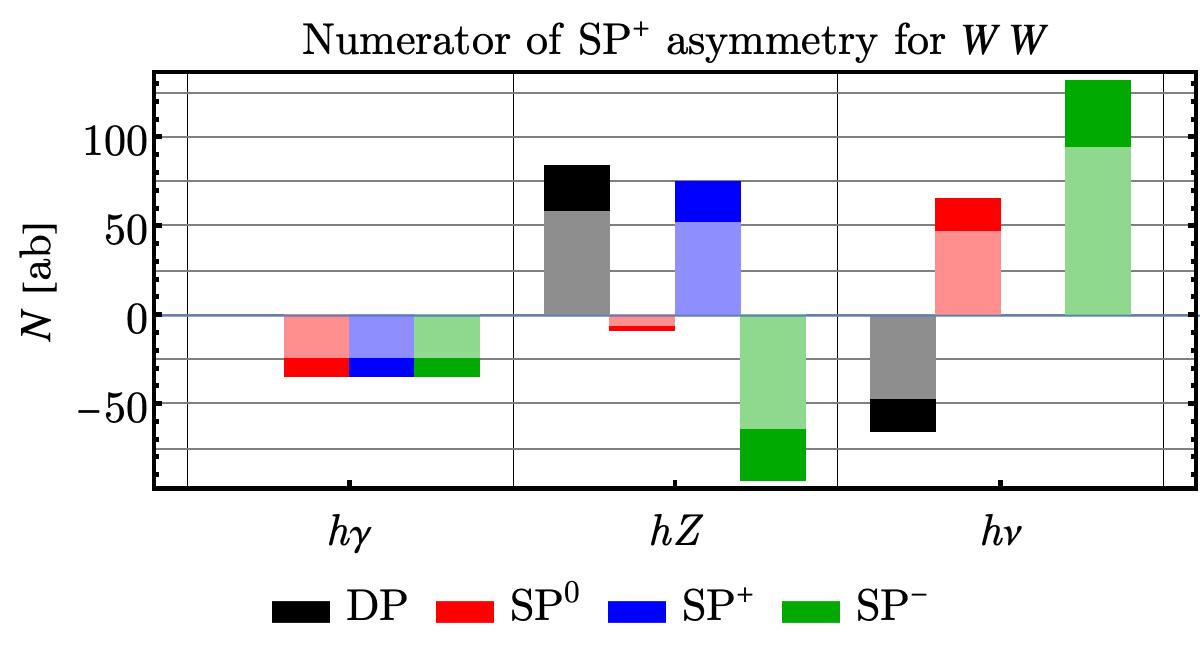}
    \includegraphics[width=.47\textwidth]{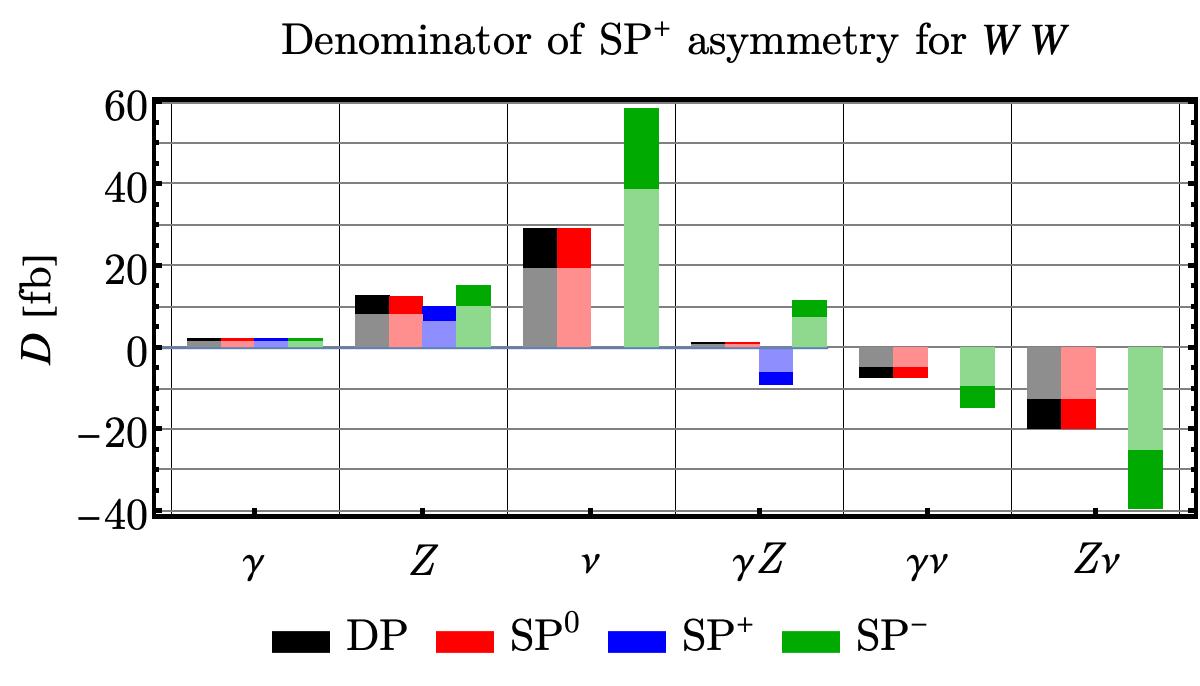}
    \caption{The split of the asymmetry numerator and denominator into their diagrammatic constituents. }
    \label{fig:WW_pchan}
\end{figure}

In this channel there is a large increase in the significance when we longitudinally polarize the positron beam. To see why this is the case we split the asymmetry numerator and denominator into their diagrammatic contributions in Fig.~\ref{fig:WW_pchan}.  The darker and lighter histograms refer to the basic and enhanced cuts, respectively. We focus first on the asymmetry numerators. We note that there is significant cancellation between the Higgs-neutrino interference and the other diagrammatic contributions. For example, for the ${\rm SP}^0$ asymmetry with basic cuts the Higgs-neutrino inteference gives 66 ab, while the other diagrammatic contributions give $-44$ ab, leading to a net asymmetry numerator of 22 ab. Due to the left-handed nature of neutrino interactions this contribution is turned off for the ${\rm SP}^+$ asymmetry, leading to a numerator of 40 ab. This larger numerator enhances the asymmetry. Similarly the denominator of ${\rm SP}^+$ is smaller due to the lack of the squared neutrino diagram, again enhancing the asymmetry. While this leads to a smaller cross section and consequently a larger statistical uncertainty, the increase of the asymmetry overwhelms this effect. These two contributions combine to give the enhancement observed for the ${\rm SP}^+$ asymmetry, and emphasize the gain from incorporating longitudinal polarization into the analysis.

\subsection{Analysis summary}

\begin{figure}[htbp]
    \centering
    \includegraphics[width=.4875\textwidth]{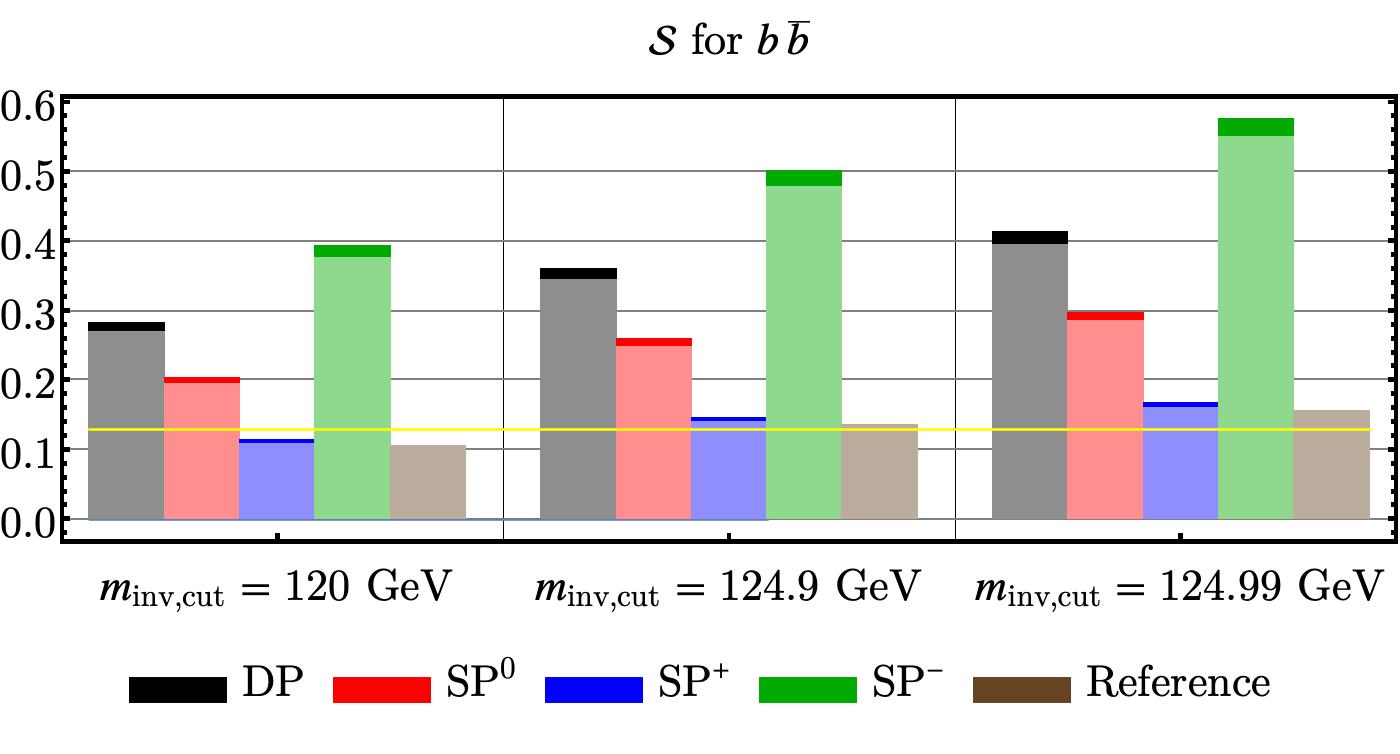}
    \includegraphics[width=.4875\textwidth]{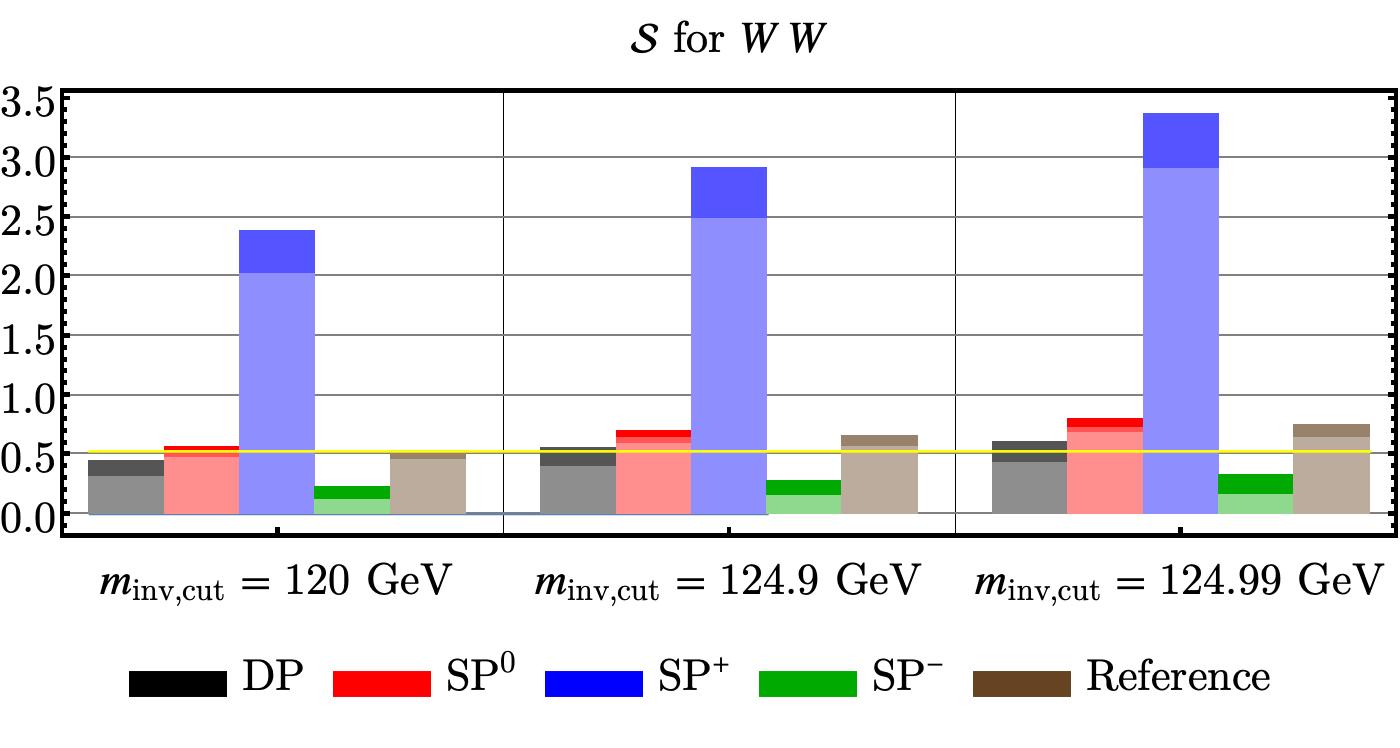}
    \caption{The significance obtained for the $b\bar{b}$ and semi-leptonic $WW$ channels for three different choices of invariant mass cut. The darker histograms refer to the optimized polar-angle cuts. }
    \label{fig:summsig}
\end{figure}

We summarize here the significance for both channels, and study the dependence of our results on the achievable beam spread and luminosity at a future FCC. The significances possible for each polarization asymmetry for three different choices of invariant mass cuts are shown in Fig.~\ref{fig:summsig}. The more stringent this cut can be made, the higher the significance that can be reached. A tighter cut increases the polarization asymmetries by reducing the continuum background and therefore the denominator of the asymmetry. The achievable cut depends on the reconstruction precision possible with future FCC detectors.

\begin{figure}[htbp]
    \centering
 \includegraphics[width=.4875\textwidth]{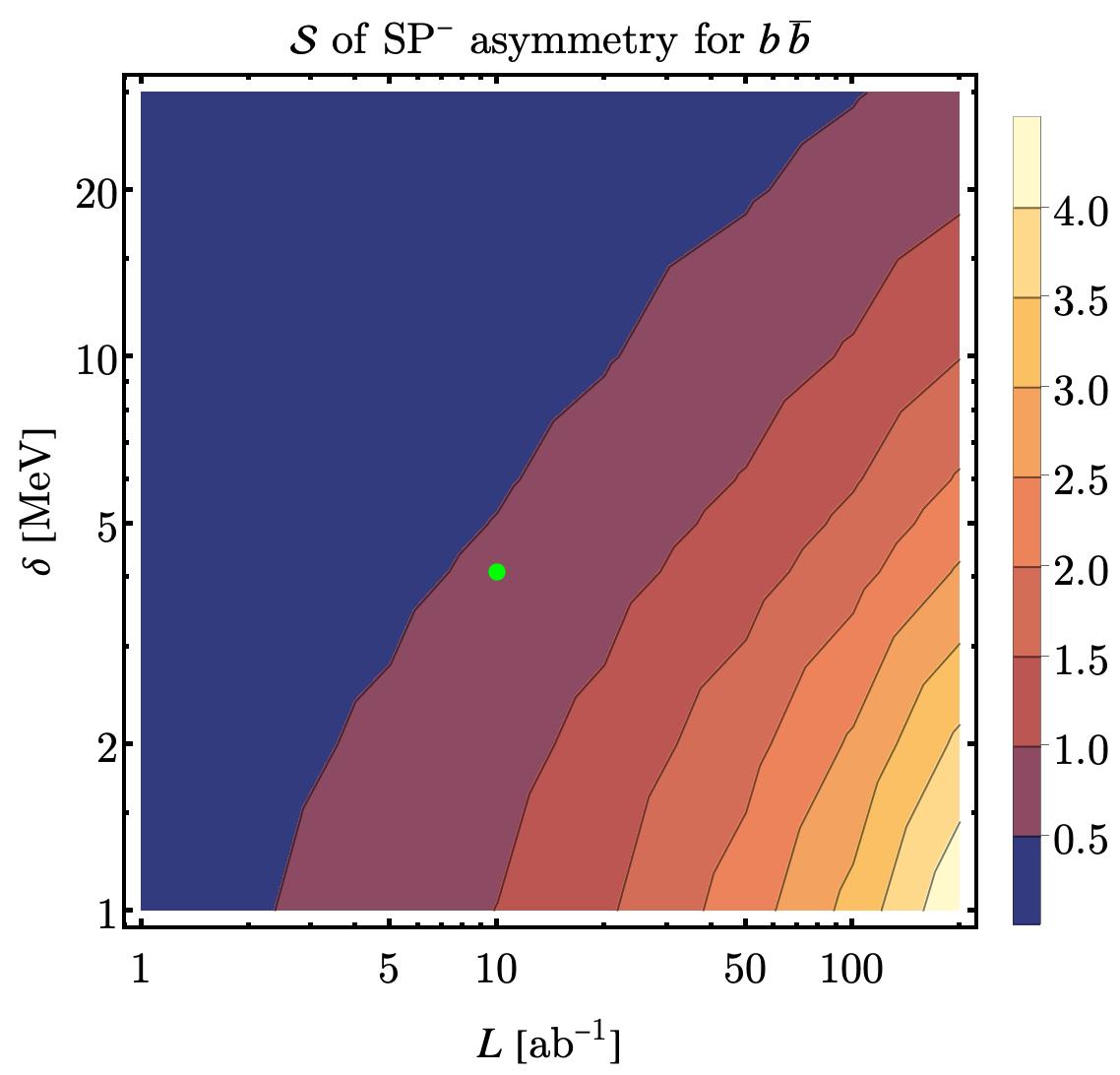}
 \includegraphics[width=.4875\textwidth]{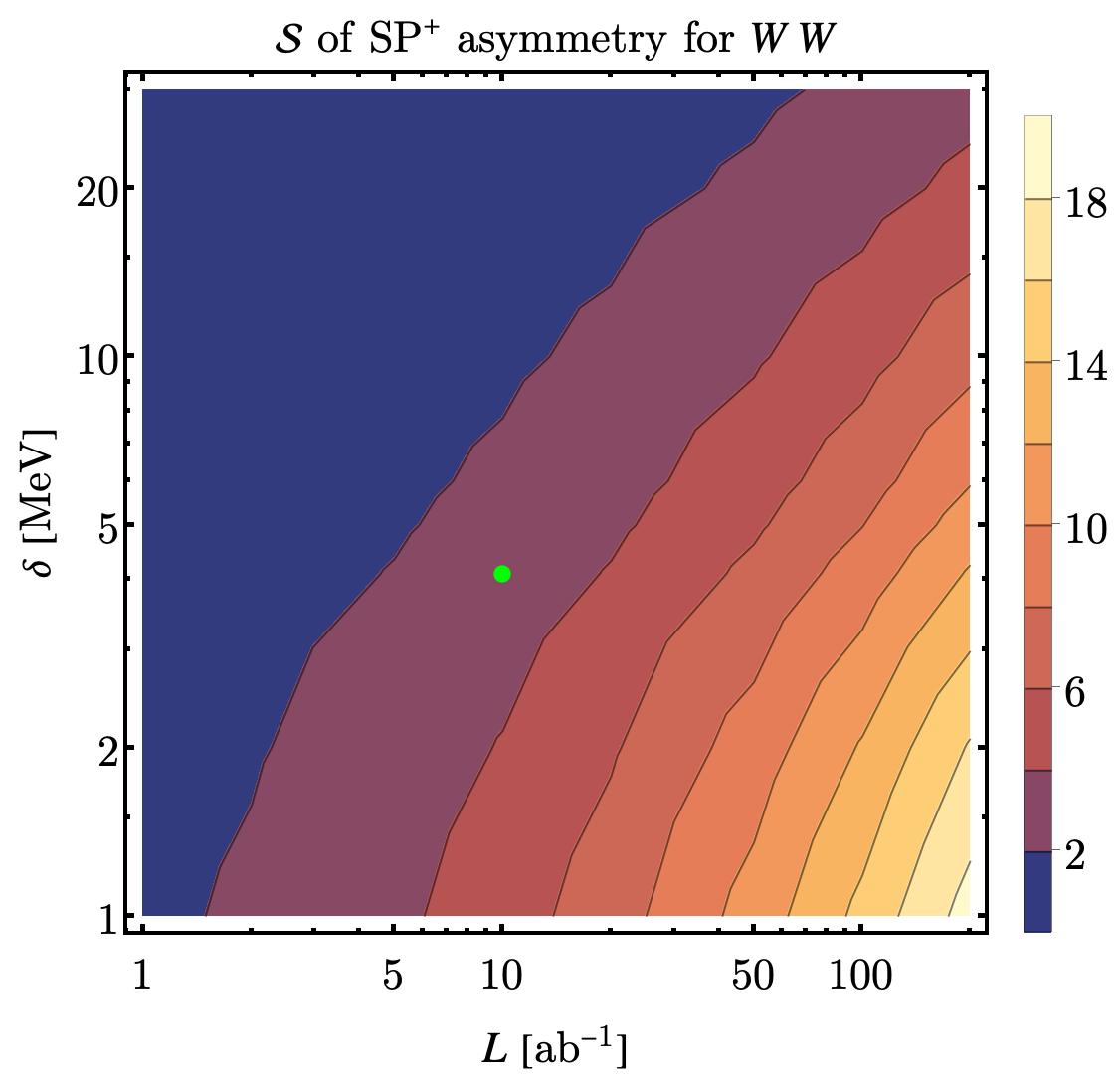}
   \caption{The significance obtained for the $b\bar{b}$ and semi-leptonic $WW$ channels as a function of the total integrated luminosity and the beam energy spread. The green dot in each panel denotes our default values $\delta = \Gamma_h$ and $L = 10 \ {\rm ab}^{-1}$}
    \label{fig:lumdep}
\end{figure}

Our analysis relies upon a future FCC complex being able to obtain both small beam energy spread and significant polarization of both beams. The first requirement is also required for the inclusive cross section determination of the electron Yukawa coupling, while the second is particular to our analysis.
It is well known that obtaining significant longitudinal polarization at an FCC is difficult, and comes at the expense of luminosity~\cite{Blondel:2019yqr}. We study the dependence on both luminosity and beam energy spread in Fig.~\ref{fig:lumdep} for both final states, for our default values of polarization  $P_{e^-} = 80\%$, $P_{e^+} = 30\%$. Since the significance only falls off as $1/\sqrt{L}$, where $L$ is the luminosity, we can still achieve $\mathcal S >2$ in the semi-leptonic $WW$ channel down to $L \approx 4 \ {\rm ab}^{-1}$ as long as the beam spread can be maintained. We can obtain $\mathcal S >1$ down to $L \approx 1 \ {\rm ab}^{-1}$. This remains a factor of two greater than the inclusive cross section reference significance obtained with  $L = 10 \ {\rm ab}^{-1}$.

\section{Conclusions}
\label{sec:conc}

We have studied how the use of single transverse-spin asymmetries can improve on the determination of the electron Yukawa coupling at a future FCC. These observables are linearly proportional to the electron Yukawa coupling since they arise from quantum interference between the Higgs signal and the continuum background, while the inclusive cross section is quadratically proportional to the electron mass. This reduces the suppression associated with this small quantity. We further study the role of longitudinal polarization of the second beam. We find that a combination of transverse polarization and even a 30\% longitudinal polarization of the second beam increases the significance in the semi-leptonic $WW$ final state to nearly three, a factor of six increase over the inclusive cross section result. Similarly the significance in the $b\bar{b}$ final state is increased by a factor of five. The primary challenge facing this measurement is the need to polarize beams while maintaining a high integrated luminosity. This is in addition to the challenges that also confront the inclusive cross section determination, which include control over beam energy spread and excellent detector resolution. We have studied the dependence on these parameters and have found that even with a luminosity reduction from 10 ab$^{-1}$ to 1 ab$^{-1}$ a significance over one is possible in the semi-leptonic $WW$ channel, still a factor of two higher than that obtained using the inclusive cross section. We encourage further studies of this possibility at a future FCC to help determine this important parameter of the SM and check whether the discovered Higgs boson indeed provides the mass of the electron.

\section*{Acknowledgments}
R.~B. and F.~P. are supported by the DOE contract DE-AC02-06CH11357.  F.~P.  and K.~S. are supported
by the DOE grant DE-FG02-91ER40684. This research was supported in part through the computational resources and staff contributions provided for the Quest high performance computing facility at Northwestern University which is jointly supported by the Office of the Provost, the Office for Research, and Northwestern University Information Technology.

\section*{Appendix}

In this section, we study the asymmetry numerator for on-shell $WW$ production. Although we use the full off-shell amplitudes in our analysis the simpler on-shell results allow us to illustrate important properties that survive in the full result. We consider
\begin{gather}
    e^-(p_a) + e^+(p_b) \to W(p_1) + W(p_2)
\end{gather}
and present analytical expressions for the $SP^+$ asymmetry numerator. We write the numerator as 
\bea
N &=& \displaystyle {1 \over 2E^2} \int d{\rm LIPS} \ \sin(\phi_{12}) \Bigg\{
    {R_{\gamma \gamma} \over s^2} 
  + {R_{ZZ} \over (s - M_Z^2)^2} 
  + {R_{H \gamma} (s - M_H^2) + I_{H \gamma} M_H \Gamma_H \over s [(s-M_H^2)^2 + M_H^2 \Gamma_H^2]} \\
  && \displaystyle + {R_{HZ} (s - M_H^2) + I_{H Z} M_H \Gamma_H \over (s - M_Z^2) [(s-M_H^2)^2 + M_H^2 \Gamma_H^2]} 
  + {R_{\gamma Z} \over s (s - M_Z^2)} 
  + {R_{\gamma \nu} \over s t}
  + {R_{Z \nu} \over t (s - M_Z^2)}
\Bigg\}
\eea

%
%
%
The $R_{ij}$ and $I_{ij}$ denote respectively the real and imaginary parts of the diagrammatic interferences between diagrams $i$ and $j$. As we assume 
the process is being studied for $s \approx M_H^2$, imaginary parts appear only for interferences involving Higgs-exchange diagrams. The real and imaginary parts can be written as
\bea
R_{\gamma\gamma} &=& \frac{e^4 m_e Q_e{}^2 (t-u) \left(-4 s M_W{}^2+12 M_W{}^4+s^2\right) p_f\cdot S_T}{M_W{}^4} \\ 
R_{ZZ} &=& \frac{e^2 m_e g_W{}^4 (t-u) \left(-4 s M_W{}^2+12 M_W{}^4+s^2\right) g_R^e \left(g_L^e+g_R^e\right) p_f\cdot S_T}{2 g_Z{}^2 M_W{}^4} \\ 
R_{H\gamma} &=& -\frac{e^2 s Q_e y_e g_W \left(12 M_W{}^4-s^2\right) p_f\cdot S_T}{2 M_W{}^3} \\
I_{H\gamma} &=& -\frac{e^2 Q_e y_e g_W \left(12 M_W{}^4-s^2\right) \epsilon \left(p_a,p_b,p_f,S_T\right)}{M_W{}^3} \\ 
R_{HZ} &=& \frac{e s y_e g_W{}^3 \left(12 M_W{}^4-s^2\right) g_R^e p_f\cdot S_T}{2 g_Z M_W{}^3} \\
I_{HZ} &=& \frac{e y_e g_W{}^3 \left(12 M_W{}^4-s^2\right) g_R^e \epsilon \left(p_a,p_b,p_f,S_T\right)}{g_Z M_W{}^3}\\ 
R_{\gamma Z} &=& -\frac{e^3 m_e Q_e g_W{}^2 (t-u) \left(-4 s M_W{}^2+12 M_W{}^4+s^2\right) \left(g_L^e+3 g_R^e\right) p_f\cdot S_T}{2 g_Z M_W{}^4} \\ 
R_{\gamma \nu} &=& \frac{e^2 m_e Q_e g_W{}^2 p_f\cdot S_T \left(M_W{}^2 \left(s^2+s u-2 t u\right)+M_W{}^4 (4 t-2 u)+s t u\right)}{2 M_W{}^4} \\
R_{Z \nu} &=& -\frac{e m_e g_W{}^4 \left(M_W{}^2-u\right) g_R^e \left(2 t M_W{}^2+4 M_W{}^4-s t\right) p_f\cdot S_T}{2 g_Z M_W{}^4}
\eea
Here we have assumed $p_f$ denotes the momentum of the leptonically-decaying $W$ boson. We note that the ${\rm sin}(\phi)$ dependence appears only in the $\epsilon \left(p_a,p_b,p_f,S_T\right)$ factor, as discussed in the main text. These terms come only with a linear factor of the Higgs Yukawa coupling $y_e$. The couplings used in this expression are:
\bea
g_Z &=& {e \over c_W s_W} \\
g_W &=& {e \over s_W} \\ 
g_L^i &=& {T_3^i - Q_i s_W^2 \over c_W s_W} \\ 
g_R^i &=& - {Q_i s_W \over c_W}.
\eea
$c_W, s_W$ denote the cosine and sine of the weak mixing angle, respectively. $e$ is the electric charge, $Q_i$ is the charge of fermion $i$ in units of $e$, and $T_3^i$ is the isospin of fermion $i$. We can form the vector and axial couplings using when discussing the $b\bar{b}$ process in the main text in terms of the left and right couplings presented here.

\bibliographystyle{h-physrev}
\bibliography{SSA}

\end{document}